\documentclass[aps,prd,superscriptaddress,nofootinbib,eqsecnum,twocolumn]{revtex4}

\pdfoutput=1

\usepackage{amsfonts}
\usepackage{amsmath}
\usepackage{amssymb}
\usepackage{graphicx,color}
\usepackage{xcolor}
\usepackage{subfigure}
\usepackage{mathtools,slashed}
\usepackage{blindtext}
\usepackage{longtable}
\usepackage{dcolumn}
\usepackage{bm}
\usepackage{appendix}
\usepackage{multirow}
\usepackage{color}
\usepackage{soul}
\usepackage{bookmark}
\usepackage{lipsum}
\usepackage{hyperref}

\begin{document}

\title{Yang-Mills extension of the Loop Quantum Gravity-corrected Maxwell equations }

\author{G. L. L. W. Levy}
\email{guslevy9@hotmail.com}
\affiliation{Centro Brasileiro de Pesquisas F\'{\i}sicas, Rua Dr. Xavier Sigaud 150, Urca, CEP 22290-180, Rio de Janeiro, RJ, Brazil}

\author{J. A. Helayël-Neto} 
\email{josehelayel@gmail.com}
\affiliation{Centro Brasileiro de Pesquisas F\'{\i}sicas, Rua Dr. Xavier Sigaud 150, Urca, CEP 22290-180, Rio de Janeiro, RJ, Brazil}

%%%%%%%%%%%%%%%%%%%%%%%%%%%%%%%%%%%%%%%%%%%%%%%%%%%%%%%%%%%%%%%%%%%%%%%%%%
\begin{abstract}

In this paper, we endeavour to build up a non-Abelian formulation to describe the self-interactions of massless vector bosons in the context of Loop Quantum Gravity (LQG). To accomplish this task, we start off from the modified Maxwell equations with the inclusion of LQG corrections and its corresponding local $U(1)$ gauge invariance. LQG effects in the electromagnetic interactions have significant importance, as they might be adopted to describe the flight time of cosmic photons coming from very high-energy explosions in the Universe, such as events of Gamma-Ray Bursts (GRBs). These photons have energy-dependent speeds, indicating that the velocity of light in the vacuum is not constant. To carry out the extension from the Abelian to the non-Abelian scenario, we shall follow the so-called Noether current procedure, which consists in recurrently introducing self-interactions into an initially free action for vector bosons by coupling the latter to the conserved currents of a global symmetry present in the action of departure. In the end of the non-Abelianization process, the initial global symmetry naturally becomes local. Once the Yang-Mills system includes LQG correction terms, it becomes possible to analyze how quantum-gravity induced contributions show up in both the electroweak and the QCD sectors of the Standard Model, providing a set-up for phenomenological investigations that may bring about new elements to discuss Physics beyond the Standard-Model.
\end{abstract}
%%%%%%%%%%%%%%%%%%%%%%%%%%%%%%%%%%%%%%%%%%%%%%%%%%%%%%%%%%%%%%%%%%%%%%

\maketitle

%%%%%%%%%%%%%%%%%%%%%%%%%%%%%%%%%%%%%%%%%%%%%%%%%%%%%%%%%%%%%%%%%%%%%%
\section{Introduction}
\label{intro}
The consistent and precise description of quantum gravity is a big challenge and an important open problem in the theory of fundamental interactions. This quest is going on for more than 50 years, by adopting different scenarios. We point out that some of those try to achieve their goal by also seeking a grand-unified theory. Loop Quantum Gravity does not enter this framework because its maajor purpose is to understand fundamental physics at the Planck scale $l_P = 10^{-35}$m, only the quantization of gravity~\cite{Ashtekar:2004eh, Ashtekar:2011ni, Rovelli:2014ssa, Rovelli:1997yv}. The LQG formulation predicts that spacetime must be discrete, granular, at the Planck scale. More precisely, the granularity of spacetime is implemented by means of a mathematically structure referred to as spin foam~\cite{Rovelli:2014ssa, Rovelli:1997yv, Thiemann:2001gmi, Gambini:2011zz}. Moreover, the fields present a background-independent description that manifests itself through an action that exhibits diffeomorphism invariance and is also non-perturbative. There are many ways to describe LQG; in the present article, we choose to work with the canonical method, which is described by Hamiltonians~\cite{Thiemann:2001gmi}. The formalism of LQG is based on the Ashtekar variables~\cite{Ashtekar:1986yd}, introduced in by 1986 Abhay Ashtekar in his formalism to describe General Relativity by an $SU(2)$ gauge connection.

Loop Quantum Gravity offers a consistent theoretical framework which extends its exploration to include phenomena that contribute to the understanding of what gravity is at a fundamental quantum scale. If we aim to detect possible traces of LQG in various events~\cite{Amelino-Camelia:1997ieq, Ellis:1999rz, Amelino-Camelia:1999hpv}, the most notable ones are: noise present in gravitational wave detectors~\cite{Amelino-Camelia:2001dbf, Amelino-Camelia:1999vks}, neutral kaon systems~\cite{Ellis:1995xd, Huet:1994kr} and the time-dependent energy of arrival of photons and neutrinos from distant sources~\cite{Alfaro:2001rb, Li:2022szn}. The electromagnetic sector has an important contribution to this. It is expected that at the Planck scale (local) Lorentz invariance, a basic principle of General Relativity, be broken. However, achieving the energy scales necessary to test both Lorentz-invariance violation (LIV) and Planck-scale effects in terrestrial experiments becomes a significant challenge, although there have been attempts, such as the investigation using of kaon systems~\cite{Amelino-Camelia:1997ieq, Li:2022szn, Gambini:1998it, Amelino-Camelia:1996bln, Melo:2024gxl}. An alternative is to focus on the energy of Gamma-Ray Bursts (GRBs) and their enormous energy emitted, which may render it possible to investigate the energy-dependent time of arrival of photons/neutrinos from distant sources. In this work, we are committed with the photon sector. The speed of light in a vacuum characterized by a granular space-time may assume the form
\begin{equation}
v(E) \simeq c\left(1- \frac{E}{E^{\gamma}_{LIV}}\right),    
\end{equation}
where $E^{\gamma}_{LIV} \gtrsim 3.6 \times 10^{17}$ GeV and is independent of the photon's helicity. The high-energy scale of the $E^{\gamma}_{LIV}$-parameter supports the idea that GRBs are the most suitable candidates to investigate this parameter ~\cite{Shao:2009bv, Zhang:2014wpb, Xu:2016zxi, Xu:2016zsa, Amelino-Camelia:2016ohi, Amelino-Camelia:2017zva, Xu:2018ien, Liu:2018qrg, Li:2020uef, Zhu:2021pml, Chen:2019avc}.

Yang-Mills (Y-M) theories are widely adopted by virtue of their effectiveness in describing the physics of fundamental interactions with the remarkable attribute of naturally opening up the path to unification. In LQG, it could not be different; there are already several works that present such a formulation as a result of the complete theory such as~\cite{Thiemann:2001gmi, Rovelli:1997yv, Gambini:1996ik}. In this paper, our goal is not to build up a non-Abelian theory in the full-fledged LQG. It is important to stress that our actual purpose is to start off from the already-known set of Maxwell's equations including LQG effects and, through the so-called the Noether's procedure to write down self-interactions among spin-1 fields, we get to the non-Abelian extension of the LQG Maxwell's equations, which we adopt the LQG-corrected Yang-Mills field equations. We intend thereby to describe a way how quantum gravitational effects, as dictated by LQG, may be brought to the Yang-Mills theory. This helps us in pursuing an investigation of the physics of self-interacting massless vector bosons in presence of tiny quantum gravity effects with the LQG signature. Effects of quantum gravity on both photons and neutrinos may be studied, as it can be seen in~\cite{Li:2022szn}. Inspired by this reference paper, our work, by putting together LQG and Yang-Mills, may offer a framework to study how the electroweak and QCD sectors get quantum-gravitational corrections, contributing to an interesting phenomenological analysis and raising the possibility of new hidden physics. The working method adopted in this paper is, as already mentioned, the so-called Noether method~\cite{Deser:1969wk,VanNieuwenhuizen:1981ae}. It is not a very popular method, but is very effective in describing this type of problem and is widely found in the literature of supersymmetry and supergravity.
 
In this paper, we are adopting the natural system of units, $c=\hbar=1$ and the signature of Minkowski metric is $(+,-,-,-)$. Our work is outlined as follows. In Sec.~\ref{sec2}, we briefly introduce the electromagnetic sector in LQG; for this, we consider the Hamiltonian of the theory and the four modified Maxwell equations in the context of Loop Quantum Gravity. In Section ~\ref{sec3}, we show the details how, from the Abelian electromagnetic sector with its Hamiltonian and equations of motion, we can obtain a non-Abelian theory of self-interactions for massless vector (gauge) bosons. {}To conclude, in Sec.~\ref{Conclusive Considerations}, we present our final comments and discussions. The Noether procedure applied to get the Yang-Mills formulation from the classical Maxwellian electromagnetic theory is summarized also given in an
Appendix~\ref{AppA}.

%%%%%%%%%%%%%%%%%%%%%%%%%%%%%%%%%%%%%%%%%%%%%%%%%%%%%%%%%%
\section{LQG Effects in the Electromagnetic Sector }
\label{sec2}
The Hamiltonian formulation of the effects of LQG in the electromagnetic theory was carried out in the work of Ref. ~\cite{Alfaro:2001rb}. This construction was done by considering the spacetime as a manifold $M$ whose topology is a diffeomorphism for $\Sigma \times \mathbb{R}$, where $\Sigma$ is a Riemmannian-3 manifold. The LQG formalism is based on the Ashtekar variables described by a non-compact gauge group, $SL(2,\mathbb{C})$, where the indices $a, b, c$ stand for the space coordinates and $i, j, k$ are their indices assigned to the generators of $su(2)$. The gauge connection is represented as $(A_a^i = \Gamma_a^i - iK_a^i)$, with $\Gamma_a^i$, torsion-free, is a function of the co-triads $e^i_a$, which in turn are defined as a function of the three-metric $q_{ab} = e^i_a e^i_b$. The term $K_a^i$ is the extrinsic curvature of the Riemmannian 3-manifold, $\Sigma$, and is related to the co-triads according to the relation $K_{ab} = sgn[\mbox{det}(e_c^j)]K_a^ i e_b^i$. The other variable part of Ashtekar's formalism is the canonical pair of the conjugate moment, namely, the electric field of the form $E_i^a = \frac{1}{2}\epsilon^{abc}\epsilon_{ijk }e_b^je_c^k$. The Hamiltonian of the electromagnetic sector modified by LQG is built up based on these considerations, and it takes the form:

%%%%%%%%%%%%%%%%%%%%%%%%%%%%%%%%%%%%%%%%%%%%%%%
\begin{align}
\label{Hamiltonian1}
H_{LQG} =& \frac{1}{Q^2} \int d^3x \Biggl \{ \left[1+\theta_7 \left (\frac{l_P}{\mathcal{L}}\right)^{2+2\Upsilon}\right]\frac{1}{2} (\underline{\Vec{B}}^2 + \underline{\Vec{E}}^2)\nonumber \\
\;+ & \theta_3 l_P^2 (\underline{B}^a \nabla^2 \underline{B}_a + \underline{E}^a \nabla ^2 \underline{E}_a) + 
\theta_2 l_P^2 \underline{E}^a \partial_a \partial_b \underline{E}^b\nonumber \\
\; + & \theta_8 l_P[\underline{\Vec{B}}\cdot  (\nabla \times \underline{\Vec{B}})+\underline{\Vec{E}}\cdot  (\nabla \times \underline{\Vec{E}})]\nonumber \\
\; + & \theta_4 \mathcal{L}^2 l_P^2 \left (\frac{\mathcal{L}}{l_p}\right)^{2\Upsilon} (\underline{\Vec{B}}^2)^2 + ...  \Biggl\},    \end{align} 
%%%%%%%%%%%%%%%%%%%%%%%%%%%%%%%%%%%%%%%%%%%%%%%
where $Q^2$ is the coupling constant of electromagnetism, $l_p \approx 1.6 \times 10^{-35}m$ is the Planck length. The characteristic length, $\mathcal{L}$, satisfies the condition $l_P \ll \mathcal{L} \le \lambda$. The parameter $\lambda$ is the de Broglie wavelength, and the characteristic length, $\mathcal{L}$, has a maximum value at momentum $k$ whenever $\mathcal{L}= k^{-1}$. Another parameter appearing in the equation is $\Upsilon$, which is given by the order of the contribution of the gravitational connection to the expected value and it may be determined by the phenomenological analysis of a selected event. That is, it may depend on the helicity of the particle under consideration~\cite{Ellis:1999uh, Ellis:1999yd, Ellis:1999sd}. The $\theta_i$'s are non-dimensional parameters of order one or are extremely close to zero~\cite{Alfaro:2001rb, Alfaro:2002ya}; finally, $a, b$ are spatial tensor indices. From now on, we replace the space indices by adopting a vector notation, and we eliminate the underline that refers to the canonical pairs for the electromagnetic sector in all quantities. From the Eq.~\eqref{Hamiltonian1}, we get the field equations as follows below:
%%%%%%%%%%%%%%%%%%%%%%%%%%%%%%%%%%%%%%%%%%%%%%%%%%%%%%%%%%
\begin{equation}\label{EM11}
\nabla \cdot \Vec{E} = 0,  
\end{equation}
%%%%%%%%%%%%%%%%%%%%%%%%%%%%%%%%%%%%%%%%%%%%%%%%%%%%%%%%%%
\begin{align}
\label{AMm2}
& A_{\gamma} (\nabla \times \Vec{B}) - \frac{\partial \Vec{E}}{\partial t} + 2 l_P^2 \theta_3 \nabla^2  (\nabla \times \Vec{B}) - 2  \theta_8 l_P \nabla^2 \Vec{B}\nonumber \\
\; & + 4 \theta_4 \mathcal{L}^2 \left (\frac{\mathcal{L}}{l_P}\right)^{2\Upsilon_{\gamma}}l_P^2 \nabla \times  (\Vec{B}^2 \Vec{B})=0,
\end{align}
%%%%%%%%%%%%%%%%%%%%%%%%%%%%%%%%%%%%%%%%%%%%%%%%%%%%%%%%%%
\begin{equation}
\label{EM22}
\nabla \cdot \Vec{B} = 0,    
\end{equation}
%%%%%%%%%%%%%%%%%%%%%%%%%%%%%%%%%%%%%%%%%%%%%%%%%%%%%%%%%%
\begin{align}
\label{FLm2}
& A_{\gamma}  (\nabla  \!\times\! \Vec{E}) \!+\! \frac{\partial \Vec{B}}{\partial t} \!+\! 2 l_P^2 \theta_3 \nabla^2  (\nabla \!\times\! \Vec{E})  \!-\!  2  \theta_8 l_P \nabla^2 \Vec{E}=0, 
\end{align}
%%%%%%%%%%%%%%%%%%%%%%%%%%%%%%%%%%%%%%%%%%%%%%%%%%%%%%%%%%
with
\begin{equation}
\label{cte}
	A_{\gamma} = 1 + \theta_7                  \left(\frac{\ell_p}{\mathcal{L}}\right)^{2+2\Upsilon}.    
\end{equation}
%%%%%%%%%%%%%%%%%%%%%%%%%%%%%%%%%%%%%%%%%%%%%%%%%%%%%%%%%%
The Eq.~\eqref{Hamiltonian1} is an infinite expression. In the literature, it is common to truncate it at the first order in the non-linear term in the magnetic field. When the field equations are calculated, this non-linear term contributes to Eq.~\eqref{AMm2}. The main quantities of the electromagnetic theory have already been calculated and analyzed previously, as it can be seen in~\cite{Alfaro:2001rb, Li:2020uef}. In the present work, such quantities are not of primary interest; the modified Maxwell equations in the context of LQG as well as its corresponding Hamiltonian are the key equations for the development of the present contribution.
%%%%%%%%%%%%%%%%%%%%%%%%%%%%%%%%%%%%%%%%%%%%%%%%%%%%%%%%%%
\section{The Noether Procedure as our path towards the non-Abelian scenario}
\label{sec3}
To formulate a non-Abelian theory for massless spin-1 particles, one may follow different approaches. In this paper, we pick out the so-called Noether method~\cite{Deser:1969wk,VanNieuwenhuizen:1981ae}, which essentially consists in generating self-interactions among vector fields through the variation of the original free Lagrangian and proposing self-interactions by coupling these vector fields with the Noether current of their initially global symmetry. To realize this program, we start off from an N-plet of vector fields (at the end of the procedure, they will become our Yang-Mills gauge bosons) in an arbitrary representation of the symmetry group as given below:
%%%%%%%%%%%%%%%%%%%%%%%%%%%%%%%%%%%%%%%%%%%%%%%%%%%%%%%%%%
\begin{equation}
\label{fields}
A_m, m=1,2,...,N  \ \ \  \ \ \ \ \ \phi_n, n=1,2,...,N
\end{equation}
%%%%%%%%%%%%%%%%%%%%%%%%%%%%%%%%%%%%%%%%%%%%%%%%%%%%%%%%%%
in an N-dimensional representation of the symmetry group. We can define  the electric- and magnetic-like fields in terms of the family of the potential field above:
\begin{equation}
\label{electricfieldpotential}
 \Vec{E}_m = -\nabla \cdot \phi_m - \frac{\partial \Vec{A}_m}{\partial t},     
\end{equation}
and
\begin{equation}
\label{magneticfieldpotential}
 \Vec{B}_m = \nabla \times \Vec{A}_m.   
\end{equation}
It is important to emphasize that the indices that shall be used from now on are associated with the N-dimensional representation of a general gauge group; they do not correspond any longer to the $su(2)$ mentioned in the previous Section. The symmetry transformations of the potential fields, Eqs.~\eqref{electricfieldpotential}, and \eqref{magneticfieldpotential} read as given below:
%%%%%%%%%%%%%%%%%%%%%%%%%%%%%%%%%%%%%%%%%%%%%%%%%%%%%%%%%%
\begin{equation}
\phi_m' = R_{mn}\phi_n,    
\end{equation}
%%%%%%%%%%%%%%%%%%%%%%%%%%%%%%%%%%%%%%%%%%%%%%%%%%%%%%%%%%
and
%%%%%%%%%%%%%%%%%%%%%%%%%%%%%%%%%%%%%%%%%%%%%%%%%%%%%%%%%%
\begin{equation}
A_m' = R_{mn} A_n.    
\end{equation}
%%%%%%%%%%%%%%%%%%%%%%%%%%%%%%%%%%%%%%%%%%%%%%%%%%%%%%%%%%
Consider an auxiliary function of the Lie group, which can be expressed in infinitesimal form as~\cite{Peskin:1995ev}:
%%%%%%%%%%%%%%%%%%%%%%%%%%%%%%%%%%%%%%%%%%%%%%%%%%%%%%%%%%
\begin{equation}
R_{mn}=(e^{iw_h G_h})_{mn} \sim \delta_{mn} + i w_h(G_h)_{mn}+ O(w^2),    
\end{equation}
%%%%%%%%%%%%%%%%%%%%%%%%%%%%%%%%%%%%%%%%%%%%%%%%%%%%%%%%%%
similarly, the fields can also be expressed in a form
\begin{equation}
\label{transformation1}
\delta A_m = i w_h(G_h)_{mn} A_n,    
\end{equation}
and
\begin{equation}
\label{transformation2}
\delta \phi_m = i w_h(G_h)_{mn} \phi_n,    
\end{equation}
note that $(G_h)_{mn}$ are the generators in the adjoint representation; they are written in terms of the structure constants as given below:
\begin{equation}
\label{generator}
[G_h,G_i] = if_{hij} G_j.    
\end{equation}
We are working in the general case of an $SU(N)$ gauge group. So, $w_h$ stands for the parameters of the $SU(N)$ transformations, whereas the 3-index symbol $f_{hij}$ represents the structure constants of $SU(N)$. To initiate the Noether procedure, we start off with the free Lagrangian of the theory. To get this Lagrangian, we perform a Legendre transformation by taking Eq.~\eqref{Hamiltonian1} and writing
\begin{align}
\label{Lagrangian}
L_{LQG} &= \frac{1}{Q^2} \int d^3x \Biggl \{  \left[1+\theta_7 \left (\frac{l_P}{\mathcal{L}}\right)^{2+2\Upsilon}\right]\frac{1}{2} (\Vec{E}^2- \Vec{B}^2)\nonumber \\
\; & +\theta_3 l_P^2 (\Vec{B} \nabla^2 \Vec{B} + \Vec{E} \nabla^2 \Vec{E}) + 
\theta_2 l_P^2 \Vec{E} \cdot [\nabla \cdot (\nabla \Vec{E})]\nonumber \\
\; & +\theta_8 l_P[\Vec{B}\cdot  (\nabla \times \Vec{B})+\Vec{E}\cdot  (\nabla \times \Vec{E})]\nonumber \\
\; & +\theta_4 \mathcal{L}^2 l_P^2 \left (\frac{\mathcal{L}}{l_p}\right)^{2\Upsilon} (\Vec{B}^2)^2 \Biggl\},    
\end{align} 
Using the variational principle, we have
\begin{equation}
\delta S = \int d^4x \ \delta L_{LQG}=0.    
\end{equation}
By considering the modified Maxwell equations in the context of LQG, it is possible to extract the (on-shell) conserved currents. Notice that these  currents carry the index ($h$) of the adjoint representation of the symmetry group, while the fields carry the index of an arbitrary representation, ($m$). Therefore, invariant self-interaction can only occur if these indices coincide, i.e., $m = h$. To allow for self-interaction between fields and currents, it is necessary that the fields, which were previously in an arbitrary representation, be placed in the adjoint representation, as the currents are. The matrix elements of the generators in the adjoint representation are written in terms of the structure constants of the symmetry group: 
\begin{equation}
\label{adjoint}
(G_h)_{ij} = -i f_{hij},    
\end{equation}
so the spatial current is
\begin{align}
\Vec{j}_h \!&=\! -f_{hij}\{A_{\gamma}[(\Vec{E}_i \cdot \phi_j) \!-\! (\Vec{B}_i \!\times\! \Vec{A}_j)]-\theta_3 l_P^2[( \Vec{B}_i \!\times\! \nabla^2 \Vec{A}_j)  \nonumber \\
\; & -(\nabla^2 \Vec{B}_i \times \Vec{A}_j)-(\nabla^2 \Vec{E}_i \cdot \phi_j)-(\Vec{E}_i \cdot \nabla^2\phi_j)]\nonumber \\
\; & -\theta_8 l_P[2(\nabla^2 \Vec{A}_i \times \Vec{A}_j)-(\Vec{E}_i \times \Vec{\dot{A}}_{j})-(\nabla \times \Vec{E}_i)\phi_j]\nonumber \\
\; & +4\theta_4 \mathcal{L}^2 l_P^2 \left (\frac{\mathcal{L}}{l_p}\right)^{2\Upsilon}(B^2 \cdot \Vec{B}_i \times \Vec{A}_j)\}, 
\end{align}
and the temporal current as
\begin{align}
j_h^0 &= -f_{hij}\{A_{\gamma}(\Vec{E}_i \cdot \Vec{A}_j)+\theta_3 l_P^2[ (\nabla^2 \Vec{E}_i \cdot \Vec{A}_j)\nonumber \\
\; & +(\Vec{E}_i \cdot \nabla^2 \Vec{A}_j)]- 2\theta_8 l_P (\nabla \times \Vec{E}_i)\cdot \Vec{A}_j\}.
\end{align}
By coupling the Noether currents to the fields , we go over from the free regime to one that exhibits self-interaction among the massless vector fields. We then introduce the coupling between the currents and the fields to get a new Lagrangian, no more free. This yields:
\begin{equation}
\label{relação lagrangiano1}
L_{LQG}^1 = L_{LQG} - l \phi_h j_h^0 - l \Vec{A_{h}}\Vec{j}_h,    
\end{equation}
the parameter $l$ is the coupling constant for the self-coupling introduced above. The development is not yet complete, because it is necessary to remove any dependence of the partial Lagrangian on space-time derivatives of the vector fields. This means that we have repeat the same procedure again. Due to the currents presenting long expressions, we chose not to present them here. Let us consider the following infinitesimal transformations for the fields
\begin{equation}
\label{transformation3}
\delta A_h = iw_{k}(G_{k})_{hl}A_l = w_{k}f_{khl}A_l,       \end{equation}
and
\begin{equation}
\label{transformation4}
\delta \phi_h = iw_{k}(G_{k})_{hl}\phi_l = w_{k}f_{khl}\phi_l.   
\end{equation}
Again, we apply the principle of least action, but now the Lagrangian used is the one in Eq.~\eqref{relação lagrangiano1}
\begin{equation}
\label{outravariação}
\delta S =  \int d^4x \ \delta L_{LQG}^1=0.   
\end{equation}
The process will be the same, but with an additional step. Before identifying the new conserved currents, we must consult the new field equations derived from the Lagrangian of the previous step. This is so because the new identified currents must be on-shell. We highlight this point, because the on-shell currents should be obey the field equations of the system at each step. From the Lagrangian, we can derive new Gauss' electricity and Ampère-Maxwell equations. This allows us to find the currents to be coupled to the fields to once again insert in the Lagrangian. Each current is coupled to a field and depends on a coupling parameter $l'$, which represents the coupling constant for this new term
\begin{equation}
\label{relação lagrangiano2}
L_{LQG}^2 = L_{LQG}^1 - l' \phi_{k} j_{k}^{0(1)} - l' \Vec{A_{k}}\Vec{j}_{k}^{(1)},    
\end{equation}
Eq.~\eqref{relação lagrangiano2} is the last Lagrangian expression obtained, and it is no longer necessary to re-calculate other currents. The reason being that the expression of the new Lagrangian no longer depends on derivatives of the fields. Should the same steps be repeated, we would obtain the same expressions for the currents as before. From Eq.~\eqref{relação lagrangiano2}, we can obtain the field equations, namely, Gauss for the electric field and Ampère-Maxwell for the magnetic field. Notice that the coupling constants $l$ and $l'$ are, in principle, arbitrary parameters of the theory. Thus, by invoking the principle of universality, we can choose them in such a way that there is only a gauge coupling constant, exactly as it is the case for Yang-Mills theories~\cite{Cheng:1984vwu}. We set $l=\frac{1}{4}g$ and $l'=\frac{1}{2}g$ and from these two equations of motion, it is possible to define the equations for the electric and magnetic fields in terms of the potentials in the Yang-Mills formalism with LQG contributions:
\begin{align}
\label{ESUN}
\Vec{E}_h \!&=\! A_{\gamma}[-\nabla  \phi_{h} \!-\! \Vec{\dot{A}}_{h} \!+\! gf_{hij}\phi_i \Vec{A}_j]\!-\! \frac{3g}{2}\theta_3 l_P^2 f_{hij}[\Vec{A}_i(\nabla^2 \phi_j) \nonumber \\
\; &  + (\nabla^2 \Vec{A}_i)\!\cdot\!  \phi_j]+\theta_3 l_P^2 f_{hij}[(\nabla \Vec{A}_i) \!\cdot\! (\nabla \phi_j)] \nonumber \\
\; & +\frac{g}{2}\theta_8 l_P f_{hij} (\Vec{A}_i \times \Vec{\dot{A}}_j),
\end{align}
and 
\begin{align}
\label{BSUN}
\Vec{B}_h &= A_{\gamma}[(\nabla \times \Vec{A}_h)+ \frac{g}{2}f_{hij}(\Vec{A}_i\times \Vec{A}_j)] + 2 l_P^2 \theta_3 \nabla^2 \Vec{B}_h \nonumber \\
\; & -\frac{g}{2} \theta_3 l_P^2 f_{hij} \cdot [\nabla^2(\Vec{A}_i\times \Vec{A}_j)- [\Vec{A}_i \times (\nabla^2 \Vec{A}_j)]]  \nonumber \\
\; & \!+\! \frac{g}{2} \theta_8 l_P f_{hij}[(\Vec{\dot{A}}_i)\!\cdot\! \phi_j] \!+\! 2\theta_4 \mathcal{L}^2 \left (\frac{\mathcal{L}}{l_P}\right)^{2\Upsilon_{\gamma}}l_P^2 [2(\Vec{B}^2 \Vec{B}_h)\nonumber \\
\; & +3g f_{hij}(\Vec{A}_i \times \Vec{A}_j)\cdot B^2].
\end{align}

We call the attention to the fact that the equations above, Eqs.~\eqref{ESUN} and \eqref{BSUN}, are the last possible ones to be obtained through our method. The procedure stops here. This is so because, if we repeat the procedure again, we will find the same currents that have been obtained in the previous step. This helps us to understand that the contributions of LQG effects in the electromagnetic sector, together with the non-Abelian formulation, add significant nonlinear terms to the general equations for the electric and magnetic fields, respectively. This is still a contribution arising from the equations of motion, which already contain these derivative terms. Here, we have all the necessary ingredients to obtain all the non-Abelian Maxwell equations, if we were working with the electromagnetism of the Standard Model of Particle Physics. As previously mentioned, the first two equations, i.e, Gauss's law for electricity and the Ampère-Maxwell equation are obtained from Eq.~\eqref{relação lagrangiano2} leaving only two equations yet to be derived. These latter equations can be obtained by taking the curl of Eq.~\eqref{ESUN} and the divergence of Eq.~\eqref{BSUN} would be sufficient. However, in electrodynamics with LQG effects, this is not possible, because the Faraday-Lenz equation is modified and, by doing this and taking the Abelian limit, we do not re-obtain the Eqs.~\eqref{EM22}, \eqref{FLm2} as it should be the case. According to the current literature~\cite{Sorokin:2021tge, Gaete:2014nda}, the non-Maxwellian extensions of electrodynamics do not modify the Faraday-Lenz equation, since together with Gauss' law for magnetism, these equations are obtained as a consequence of the Bianchi Identities written in terms of the gauge-covariant derivatives. The electrodynamics presented in this paper modifies Eq.~\eqref{FLm2}, which allows us to conclude that the Bianchi Identities are modified. It is therefore necessary to follow another path to attain these two equations; one possible way consists in getting them from the Hamilton-Jacobi equations. By performing a Legendre transformation on the Lagrangian Eq.~\eqref{relação lagrangiano2} to find its corresponding Hamiltonian, we can derive the equations for the electric field, $E$, and the vector potential, $A$:
\begin{equation}
\label{eomH}
\frac{\partial H_{LQG}^2}{\partial \Vec{E}_h}= - \frac{\partial \Vec{A}_h}{\partial t},   
\end{equation}
and 
\begin{equation}
\label{eomH2}
\frac{\partial H_{LQG}^2}{\partial \Vec{A}_h} + \frac{\partial H_{LQG}^2}{\partial(\nabla \times \Vec{A}_{h})} =  \frac{\partial \Vec{E}_h}{\partial t}.    
\end{equation}
From Eq.~\eqref{eomH} and by taking its curl, it is possible to derive the Faraday-Lenz equation. Meanwhile, Gauss's law for magnetism is obtained from Eq.~\eqref{eomH2} by taking the divergence of the magnetic field. This yields four Maxwell-type equations in the non-Abelian version including the LQG contributions :
\begin{widetext}
  \begin{align}\label{M1}
& A_{\gamma}[\nabla \cdot \Vec{E}_h + gf_{hij} \Vec{E}_i \cdot \Vec{A}_j] + \frac{g}{2}\theta_3 l_P^2f_{hij}[2(\nabla^2 \Vec{E}_i)\cdot \Vec{A}_j + \Vec{E}_i \cdot (\nabla^2 \Vec{A}_j)+\nabla^2(\Vec{A}_i\cdot \Vec{E}_j) \nonumber \\
\; & - 2\nabla[(\nabla \Vec{A}_i) \cdot (\nabla \phi_j)]]+ \frac{g}{2}\theta_8 l_P f_{hij} [\nabla \cdot (\Vec{A}_i \times \vec{\dot{A}}_j)+\Vec{A}_i \cdot (\nabla \times \Vec{E}_j)]  = 0,
\end{align}
\end{widetext}

\begin{widetext}
\begin{align}
& A_{\gamma}[\nabla \times \Vec{B}_h + gf_{hij}\phi_i \Vec{E}_j + gf_{hij}(A_i \times B_j)] + \theta_3 l_P^2\{2\nabla^2 (\nabla \times \Vec{B}_h)-gf_{hij}[(\nabla^2 \Vec{E}_i) \phi_j +\frac{1}{2}\nabla^2 (\phi_i \Vec{E}_j) \nonumber \\
\; &  + [(\nabla^2 \Vec{B}_i) \times \Vec{A}_j]- \frac{1}{2} \phi_i \nabla^2(\Vec{\dot{A}}_j)+\frac{1}{2} \Vec{E}_i (\nabla^2 \phi_j) - \frac{1}{2} (\Vec{\dot{A}}_i)\cdot (\nabla^2 \phi_j)- \frac{1}{2}\nabla^2 (\Vec{B}_i \times \Vec{A}_j)- \frac{1}{2}[\Vec{B}_i \times (\nabla^2 \Vec{A}_j)]\nonumber \\
\; & - \nabla^2 [\phi_i (\Vec{\dot{A}}_j)] ]\}  -\theta_8 l_P\{2\nabla^2 \Vec{B}_h-\frac{g}{2}f_{hij}[\partial_t(\Vec{A}_i \times \Vec{A}_j)-[(\Vec{\dot{E}}_i) \times \Vec{A}_j]-2\nabla^2(\Vec{A}_i\times \Vec{A}_j)- \nabla[\nabla \cdot (\Vec{A}_i\times \Vec{A}_j)]\nonumber \\
\; & + (\nabla \times \Vec{E}_i)\phi_j]\}
+ 2\theta_4 \mathcal{L}^2 \left (\frac{\mathcal{L}}{l_P}\right)^{2\Upsilon_{\gamma}}l_P^2 \nabla \times \left[2(\Vec{B}^2 \Vec{B}_h)+3g f_{hij}(\Vec{A}_i \times \Vec{A}_j)\cdot B^2 \right]  = \frac{\partial \Vec{E}_h}{\partial t},
\label{M2}
\end{align}
\end{widetext}

\begin{widetext}
\begin{align}
& A_{\gamma}[(\nabla \cdot \Vec{B}_h)+ gf_{hij}\Vec{A}_i \cdot \Vec{B}_j]-\frac{g}{2} \theta_3 l_P^2 f_{hij}\nabla \cdot [\vec{A}_i \times (\nabla^2 \vec{A}_j)-3\nabla^2 (\vec{B}_i \times \vec{A}_j)]+\frac{g}{2}\theta_8 l_Pf_{hij} \nabla \cdot[ (\Vec{\dot{A}}_{i}) \cdot \phi_j]\nonumber \\
\;& + 9g\theta_4 \mathcal{L}^2 \left (\frac{\mathcal{L}}{l_P}\right)^{2\Upsilon_{\gamma}}l_P^2f_{hij} \nabla \cdot [ (\Vec{A}_i \times \Vec{A}_j)\cdot B^2]=0,
\label{M3}
\end{align}    
\end{widetext}
\begin{widetext}
\begin{align}
& A_{\gamma}[(\nabla \times \Vec{E}_h)-gf_{hij}\phi_i \Vec{B}_j + gf_{hij}(\Vec{A}_i\times \Vec{E}_j)]+\theta_3 l_P^2 \nabla^2 (\nabla \times \Vec{E}_h)-\frac{g}{2}\theta_3 l_P^2f_{hij}[\Vec{B}_i \cdot (\nabla^2 \phi_j) \nonumber \\
\; & -(\nabla^2 \Vec{E}_i \times \Vec{A}_j)]-\theta_8 l_P[\nabla^2 \Vec{E}_h+ gf_{hij}\nabla \cdot (\Vec{E}_i \cdot \Vec{A}_j)-\frac{g}{2}f_{hij}\nabla \times (\Vec{A}_i\times (\Vec{\dot{A}}_{j}))]=-\frac{\partial \Vec{B}_h}{\partial t}.  
\label{M4}
\end{align}
\end{widetext}
When considering the four non-Abelian Maxwell-type equations for massless spin-1 vector bosons in the context of LQG effects, it is important to notice that when the generators of the group $SU(N)$ become trivial ($G_h = 0$), the structure constants of the group are taken $f_{hij} = 0$, we re-obtains the LQG-corrected Maxwell equations, \eqref{EM11}, \eqref{AMm2}, \eqref{EM22}, and \eqref{FLm2}. Each term in $\theta^i$ contributes independently to the structure constants, and the term $\theta_4$, coupled with the nonlinear magnetic field, modifies the new magnetic field equations. Previously, the Eq. \eqref{M3}, known as Gauss Magnetism, did not include magnetic monopoles. However, this is no longer true in the non-Abelian formulation, as terms $\theta_3$, $\theta_8$, and $\theta_4$ independently contribute to this equation. It is conceivable, based on this interpretation, that magnetic monopoles might exist at the Planck scale, possibly associated to these parameters.
%%%%%%%%%%%%%%%%%%%%%%%%%%%%%%%%%%%%%%%%%%%%%%%%%%%
\section{Conclusive Considerations}
\label{Conclusive Considerations}
In this work, all our efforts consist in formulating a non-Abelian action for massless spin-1 fields in the context of Loop Quantum Gravity. We know that Yang-Mills gauge theories give a very complete  description of fundamental interactions: this is the case of QCD and the unified electroweak model based on the $SU(2) \times U(1)$-symmetry.

To formulate the theory, we first transformed the free Hamiltonian into a free Lagrangian by means of a Legendre transformation. By applying the variational principle, we were able to obtain the time and space components of the Noether currents. Next, we introduced these currents coupled to the vector fields into the free Lagrangian to generate self-interactions. The procedure was recursively repeated until we obtained another self-interaction current, which, when coupled to the vector fields and inserted into the step-by-step corrected Lagrangian stopped the process. Additionally, by deriving new motion equations from the new Lagrangian, we obtained the Gauss Electricity and Ampère-Maxwell equations. Furthermore, we were able to define the Electric and Magnetic fields in $SU(N)$. However, unlike the Abelian Maxwell equations, we could not obtain the last two equations, namely, Gauss magnetic and Faraday-Lenz equations. To resolve this, we obtained a Hamiltonian from the final Lagrangian once again, and from the Hamilton-Jacobi motion equations, we derived the last two equations.

The LQG-modified Maxwell and Yang-Mills equations open up new phenomenological prospects, as the dimensionless parameters of the LQG contribute to these equations and introduce tiny effects into Standard-Model processes. In the Abelian case, we highlight the presence os a non-linear contribution quadratic in the magnetic field. Furthermore, this non-linear term, along with the LQG and other parameters is present in the magnetic monopole equation. So, it is possible that at the Planck scale, these parameters may yield magnetic monopole solutions. 

It is important to remark that we have not worked in the covariant formalism. The idea is to express in a manifest way field strengths and the gauge potentials interact. The developments proposed in this paper may open up the way for future developments of LQG effects in unexplored areas of physics, such as the investigation of LQG effects in both the electroweak theory and QCD. It is important to notice that effects arising from quantum gravitation may be found in these theories at high-energy scales. We are already pursuing the investigation of the effect of LQG corrections in connection with the electroweak sector of the Standard Model. More specifically, to our sense, it would be relevant to study the influence of LQG corrections on the anomalous vertices coupling the neutral gauge bosons, namely the photon and the Z0-boson. The present ATLAS and CMS experimental data could be used to get a new set of constraints on the LQG parameters. In particular, considering the electroweak sector, as mentioned, after the phase transition induced by the Higgs dublet, the third component of the SU(2) gauge boson and the weak hypercharge (Abelian) field get mixed. By performing the orthogonal transformation parametrized by the Weinberg angle, the photon and the Z0-boson field emerge as the physical fields, corresponding, respectively, to the zero mass and massive eigenstates of the mass matrix. Therefore, by introducing LQG effects in both the SU(2) and the U(1) factors, there will appear anomalous 3- and 4-vertices coupling directly the photon and the Z0-gauge boson. These vertices are searched for in the ATLAS and CMS Collaborations. Their present results may be used to establish a new bridge connecting LQG to accelerator physics.

%%%%%%%%%%%%%%%%%%%%%%%%%%%%%%%%%%%%%%%%%%%%%%%%%%%%%%
\section{Acknowledgements}

G.L.L.W.L. acknowledges financial support of the \\
Fundação Carlos Chagas Filho de Amparo à Pesquisa \\
do Estado do Rio de Janeiro (FAPERJ), Grant No. E- \\
26/202.437/2024. The authors are also grateful to P. A. Lima \\
Mourão for valuable discussions on the LQG-corrected \\
Maxwell equations in presence of background magnetic fields and astrophysical processes involving photons and neutrinos.

\appendix
\section{From Maxwell to Yang-Mills: a glance at the Noether current procedure}
Let us build up the Yang-Mills theory as a self-interacting theory of massless (gauge) vector bosons, starting from the ordinary Maxwell equations, as follows:
\label{AppA}
\begin{equation}
\label{equação (1)}
\nabla \cdot \Vec{E} = 0,    
\end{equation}
\begin{equation}
\label{equação (2)}
\nabla \cdot \Vec{B} = 0,
\end{equation}
\begin{equation}
\label{equação (3)}
\nabla \times \Vec{E} = -\frac{\partial \Vec{B}}{\partial t},    
\end{equation}
\begin{equation}
\label{equação (4)}
\nabla \times \Vec{B} =\frac{\partial \Vec{E}}{\partial t},   
\end{equation}
The fields are represented in the the same form as in Eqs.~\eqref{transformation1}, \eqref{transformation2} and \eqref{generator}. Let us start off from the free lagrangian of the electromagnetic theory:
\begin{equation}
\label{equação (16)}
\mathcal{L}_{EM} = \frac{1}{2}(E^2 - B^2).    
\end{equation}
Using the variational principle,  it is possible to find both the time and space components of the conserved currents. The space component is given by
\begin{equation}
\label{equação (22)}
\Vec{j}_h = -f_{hij} [(\Vec{E_i} \cdot \phi_j) - (\Vec{B_i} \times \Vec{A_j})],  
\end{equation}
whereas the time component reads as
\begin{equation}
\label{equação (23)}
j_h^0 = -f_{hij} (\Vec{E_i} \cdot \Vec{A_j}).  \end{equation}
Notice that the currents transform under the adjoint representation of the gauge group. So, they have a representation index different from the indices carried by the fields, since the latter are sitting in a general N-dimensional representation. To couple the currents to the potential fields in a gauge-invariant way, the two indices have to match. So, the gauge potentials must necessarily be put in the adjoint representation. In so doing, we can immediately couple the currents to the fields to get a new Lagrangian. Important: after the current/vector field coupling, the Lagrangian is no longer free; now, there appear self-interactions.
\begin{equation}
\label{coupledlagrangian}
\mathcal{L}_{EM}^1 = \frac{1}{2}(E^2 - B^2) - l \phi_h j_h^0 - l \Vec{A_h}\Vec{j_h},    
\end{equation}
where $l$ is the free coupling constant parameter. We still have the dependence on space-time derivatives of the vector fields. To remove them, the field transformations are Eqs.~\eqref{transformation3}, \eqref{transformation4}. From the Eq.~\eqref{coupledlagrangian}, it is possible calculate the new field equations and the new on-shell currents associated to the Lagrangian:
\begin{equation}
\label{equação (30)}
\Vec{j_{k}^{(1)}}=-f_{khl}[(\Vec{E}_h \cdot \phi_l)-(\Vec{B}_h \times \Vec{A}_l)-2lf_{hij}\Vec{A}_i\phi_l\phi_j],    
\end{equation}
\begin{equation}
\label{equação (31)}
j_{k}^{0(1)}=-f_{khl}( \Vec{E}_h \cdot \Vec{A}_l - 2lf_{hij}\Vec{A}_i\Vec{A}_l\phi_j ).    
\end{equation}
These two currents above signal the final step of ther procedure; so, we can couple them through the Lagrangian
\begin{equation}
\label{coupledlagrangian2}
\mathcal{L}_{EM}^2 = \mathcal{L}_{EM}^1 - l' \phi_{k} j_{k}^{0(1)} - l' \Vec{A_{k}}\Vec{j}_{k}^{(1)},    
\end{equation}
Again, the $l'$ is another free coupling constant parameter. From the Eq.~\eqref{coupledlagrangian2}, it is possible obtain the equation of the electric and magnetic field in the Yang-Mills formalism
\begin{equation}
\label{electricsun}
\Vec{E}_{h} = -\nabla \cdot \phi_{h} - \frac{\partial \Vec{A}_{h}}{\partial t} + gf_{hij}\Vec{A}_i \phi_j,    
\end{equation}
\begin{equation}
\label{magneticsun}
\Vec{B}_{h} = (\nabla \times \Vec{A}_{h})+ \frac{1}{2}gf_{hij}(\Vec{A}_i \times \Vec{A}_j).  
\end{equation}
\\
It is possible to obtain the Gauss' electricity and Ampère-Maxwell equations from Eq.~\eqref{coupledlagrangian2}. However another strategy consists of obtaining the four non-Abelian Maxwell equations from the general form of the field Eqs.~\eqref{electricsun} and \eqref{magneticsun}. If we take the curl and div from both equations, we can obtain the new "non-Abelian" Maxwell equations: actually, the Yang-Mills equations.
\begin{equation}
\label{equação (40)}
\nabla \cdot \Vec{E}_{h} + g f_{hij}\Vec{A}_i \Vec{E}_j = 0,    
\end{equation}
\begin{equation}
\label{equação (41)}
\nabla \cdot \Vec{B}_{h} + gf_{hij}\Vec{A}_i \Vec{B}_j=0,    
\end{equation}
\begin{equation}
\label{equação (42)}
\nabla \times \Vec{E}_{h} + g f_{hij} (\Vec{A}_i \times \Vec{E}_j) = - \frac{\partial \Vec{B}_{h}}{\partial t}  +  g f_{hij} \phi_i \Vec{B}_j,   
\end{equation}
\begin{equation}
\label{equação (43)}
\nabla \times \Vec{B}_{h} + g f_{hij}(\Vec{A}_i \times \Vec{B}_j) = \frac{\partial \Vec{E}_{h}}{\partial t} - gf_{hij} \phi_i \Vec{E}_j. 
\end{equation}

%%%%%%%%%%%%%%%%%%%%%%%%%%%%%%%%%%%%%%%%%%%%%%%%%%%%%%

\end{document}